\documentclass[osajnl,twocolumn,showpacs,superscriptaddress,10pt]{revtex4-1} 
\usepackage{amsmath,amssymb,graphicx}
\begin{document}


\title{Efficiency of dispersive wave generation in dual concentric core microstructured fiber}

\author{D.~Modotto}\email{Corresponding author: daniele.modotto@unibs.it}
\affiliation{Dipartimento di Ingegneria dell'Informazione, Universit\`a di Brescia, via Branze 38, 25123 Brescia, Italy}
\author{M.~Andreana}
\affiliation{Department of Physics, University of Ottawa, Ottawa ON, Canada}
\author{K.~Krupa}
\affiliation{XLIM, Universit\'e de Limoges, UMR CNRS 7252, 123 Av. A. Thomas, 87060 Limoges, France}
\author{G.~Manili}
\author{U.~Minoni}
\affiliation{Dipartimento di Ingegneria dell'Informazione, Universit\`a di Brescia, via Branze 38, 25123 Brescia, Italy}
\author{A.~Tonello}
\author{V.~Couderc}
\author{A.~Barth\'el\'emy}
\author{A.~Labruy\`ere}
\affiliation{XLIM, Universit\'e de Limoges, UMR CNRS 7252, 123 Av. A. Thomas, 87060 Limoges, France}
\author{B.~M.~Shalaby}
\affiliation{XLIM, Universit\'e de Limoges, UMR CNRS 7252, 123 Av. A. Thomas, 87060 Limoges, France}
\affiliation{Physics Department, Faculty of Science, Tanta University, Tanta, Egypt}
\author{P.~Leproux}
\affiliation{XLIM, Universit\'e de Limoges, UMR CNRS 7252, 123 Av. A. Thomas, 87060 Limoges, France}
\author{S.~Wabnitz}
\affiliation{Dipartimento di Ingegneria dell'Informazione, Universit\`a di Brescia, via Branze 38, 25123 Brescia, Italy}
\author{A. B. Aceves}
\affiliation{Department of Mathematics, Southern Methodist University, Dallas, USA}

\begin{abstract}
We describe the generation  of powerful dispersive waves that are observed when pumping a dual concentric core  microstructured 
fiber by means of a sub-nanosecond laser emitting at the wavelength of~1064 nm. 
The presence of three zeros in the dispersion curve, their spectral separation from the pump wavelength, and
the complex dynamics of solitons originated by the pump pulse break-up, all contribute to boost the amplitude of the dispersive wave on the long-wavelength side of the pump. The measured conversion efficiency towards the dispersive wave at 1548 nm is as high as 50\%. 
Our experimental analysis of the output spectra is completed by the acquisition of the time delays of the different spectral components. Numerical simulations and an analytical perturbative analysis identify 
the central  wavelength of the red-shifted pump solitons and the dispersion profile of the fiber as the key parameters for determining the efficiency of the dispersive wave generation process.
\end{abstract}

\ocis{(190.4370) Nonlinear optics, fibers; (190.5530) Pulse propagation and temporal solitons; (320.6629) Supercontinuum generation; (060.4005) Microstructured fibers.}

\maketitle 

\section{Introduction}

Dispersive waves (DWs), also known as Cherenkov or nonsolitonic radiations, are generated in fibers when optical soliton pulses are perturbed by higher-order dispersion terms. As initially predicted by Menyuk and co-workers for standard fibers, an intense pulse whose input spectrum is close to the zero-dispersion wavelength (ZDW) and in the presence of third-order dispersion (TOD) sheds power in the form of a long pulse, whose spectrum is a narrow peak \cite{Wai1986,Wai1987}. By assuming that the TOD and the fourth-order dispersion coefficients are known, 
and by relying on an accurate formula for the perturbed soliton, Akhmediev and Karlsson have shown that it is possible not only to calculate the DW  frequency, but also its temporal shape and the amount of radiated energy \cite{Akhmediev1995}. Even when the full dispersion curve is taken into account, the DW wavelength can be readily calculated by imposing a matching condition between the phases of the soliton and the DW. In fact it can be shown that, depending on the shape of the dispersion curve, the DW may be generated at shorter or longer wavelengths than that of the pump. Moreover, two or more DWs can be simultaneously generated \cite{Roy2009,Roy2010,Stark2011}. 

Microstructured fibers (MFs) allow for a great freedom in the design of the dispersion curve and mode effective area \cite{Eggleton2001,Russell2006,Poli2010}: the former feature is essential to control the DW spectral position and the latter helps to maximize the nonlinear coefficient. In fact, many groups have reported an efficient DW generation (for instance: \cite{Tartara2003,Chang2010,Yuan2011}) when the MF is pumped by pulses whose duration is of a few hundreds femtoseconds or even shorter. When sufficiently intense input pulses are used, the DW generation is accompanied by the formation of a supercontinuum (SC) spectrum, whose bandwidth can reach the extension of a full octave \cite{Genty2004,Travers2008,Kudlinski2009}. Different phenomena contribute to SC generation: self-phase modulation (SPM) and modulation instability (MI) lead to the initial spectral broadening, which can be further enlarged by four-wave mixing (FWM) \cite{Wadsworth2004,Stone2008,Lesvigne2007,Manili2011}, soliton fission \cite{Herrmann2002,Demircan2007,Driben2013} and Raman soliton self-frequency shift \cite{Mitschke1986}. Moreover, the large number of pulses fostered  by the break-up of a long pump pulse can also interact with the DW, leading to the onset of new spectral peaks and contributing to determine the extension and flatness of the output spectrum \cite{Genty2004bis,Skryabin2005,Gorbach2006,Driben2010}.

The development of practical DW sources is generally limited by the small efficiency of the energy transfer from the pump to the DW. In most of experiments reported to date, the power carried by the DW is only a small percentage of the input pump power; it is possible to increase the conversion efficiency (up to 65\%  in recent experiments) only by resorting to a femtosecond laser source (like a Ti:sapphire laser) \cite{Tartara2003,Chang2010,Yuan2011,Zhang2013,Zhang2015}. In this case, intense DWs have been observed in the visible range (even in the violet \cite{Zhang2013bis}) as well in the near and mid-infrared \cite{Yuan2013,Zhang2013}. 
Indeed with ultrashort pump pulses their spectrum may be so wide that it incorporates the resonant condition for DW generation.

There is a long way ahead in order to achieve high conversion efficiencies when using nanosecond pulses emitted by low-cost, widely used sources, such as Nd:YAG microchip lasers \cite{Wadsworth2004}. Nanosecond pulses cannot directly generate DWs, owing to their extremely narrowband spectra. However
high energy nanosecond pulses may break, after a first stage of nonlinear propagation in a MF, into a bunch of ultrashort pulses, which in turn can generate DWs. 

It must be emphasized that a pulse that sheds light into a DW is not necessarily restricted to a fundamental or higher-order soliton. In fact, it has been shown that intense pulses traveling in the normal dispersion regime of the fiber and satisfying a phase-matching relation may also effectively build up a DW \cite{Roy2010,Webb2013}. 
Moreover, DWs have also been observed in a line-defect photonic crystal waveguide \cite{Colman2012} with a length of only 1.5~mm: the measured 30\% conversion efficiency could be explained by means of the locking of the velocities of the pump soliton and the DW.

We have recently reported the experimental observation of a gigantic DW, generated inside a dual concentric core MF pumped by a microchip laser, and carrying up to 50\% of the input pump power at the fiber output \cite{Manili2012}. In the present work, we further clarify, by means of an extensive experimental analysis, the physical mechanism for such huge energy transfer into the DW. Section 2 gives an account of the spectra measured when pumping at the wavelengths of 1064~nm and 1030 nm. Whereas Section 3 describes the spectro-temporal analysis of a 200 nm wide bandwidth centered around the emitted DW. Section 4 is devoted to the numerical  and analytical study of the DWs: it also unveils the roles of the pump soliton central wavelength
and of the fiber dispersion profile. Finally, Section 5 briefly summarizes the results of our study.

\section{Experiments with two different laser sources}

\begin{figure}[htbp]
\centerline{\includegraphics[width=.8\columnwidth]{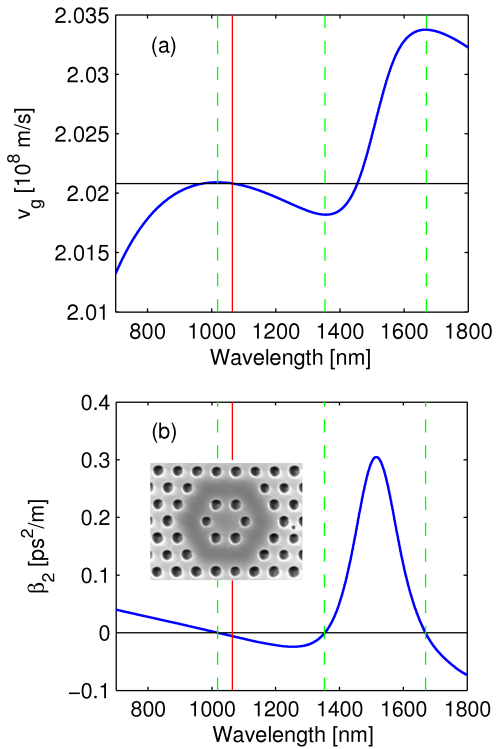}}
\caption{Numerically calculated parameters of the double core MF: (a) group velocity and (b) its dispersion. The vertical solid red line indicates the 1064~nm wavelength; vertical dashed green lines indicate the three ZDWs (1018~nm, 1353~nm, 1669~nm). The inset shows a SEM photograph of the fiber cross section.}
\label{fig1}
\end{figure}

Our dual concentric core MF has an inner core and a second external concentric annular core. These two cores are obtained by filling some of the holes in a triangular lattice; the holes have a radius of 0.65~$\mu m$ and are separated by a pitch of 2.6~$\mu m$. A scanning electron microscope (SEM) image of the fiber cross section is displayed in the inset of Fig.~\ref{fig1}: the two glass hexagonal cores are easily recognized. The central core is also doped by Germanium in order to increase both local refractive index and nonlinearity \cite{Nakajima2002,Yatsenko2009,Labruyere2010}. This kind of double core structure has been proposed and demonstrated to be a very effective design to control the magnitude and sign of the group velocity dispersion (GVD) in a selected spectral range \cite{Gerome2004,Gerome2006}.

The linear guiding properties of our double core MF can be calculated from the SEM picture through a numerical mode solver: Fig.~\ref{fig1}(a) and Fig.~\ref{fig1}(b) show the group velocity ($1/\beta_1$) and its dispersion ($\beta_2$), respectively. Looking at the GVD it is interesting to observe the presence of a large positive dispersion peak: as a result, wavelengths in the spectral region around 1650~nm are expected to overtake 
spectral components at 1450~nm by 31.5~ps/m. This condition is quite unusual, since in conventional solid core microstructured fibers wavelengths around 1650~nm are in the anomalous dispersion region. Hence typical SC generation exhibits wavelength components around 1650~nm that appear in the trailing edge of the output pulse. According to our numerical results, the normal dispersion peak reaches its maximum amplitude at 1515~nm, and it is bounded by two ZDWs at 1353~nm and at 1669~nm, respectively. It is worth noticing that in our fiber dispersion is anomalous in the range between 1018~nm and 1353~nm: as a consequence, solitons generated by the break-up of a pulse centered at 1064~nm (or at 1030~nm) may only exist in this wavelength range.
{ The fundamental mode of our MF is well confined inside the central core for wavelengths shorter than 1400~nm. Whereas for wavelengths greater than 1400~nm an important fraction of the optical power is confined by the external hexagonal core (as for the fundamental supermode of the dual concentric core fiber discussed in Ref.~\cite{Gerome2006}). It is this abrupt change of the mode profile and consequently of its dispersion relation which gives rise to the dispersion peak. For instance, the calculated effective area at 1064~nm is 6.2~$\mu m^2$ and it increases up to 41.6~$\mu m^2$ at 1550~nm. These numerical results were experimentally confirmed by inspecting the mode profile at the fiber output by means of an infrared camera.}

\begin{figure}[htbp]
\centerline{\includegraphics[width=.8\columnwidth]{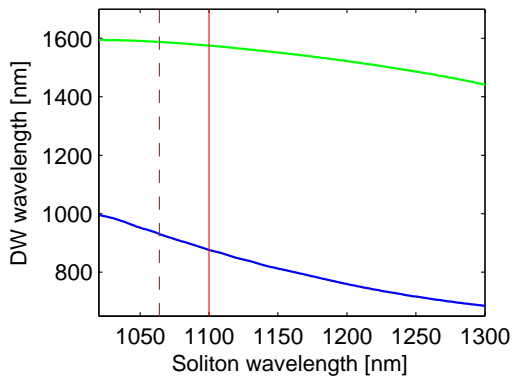}}
\caption{Calculated positions of $DW_1$ (blue line) and $DW_2$ (green line) versus the pump soliton wavelength. The vertical lines indicate the wavelengths of 1064~nm (dashed red) and 1100~nm (solid red).}
\label{fig2}
\end{figure}

By inserting the numerically calculated effective refractive index profile in the well-known soliton-DW phase-matching relation (see for instance Refs.~\cite{Akhmediev1995,Genty2004,Chang2010,Zhang2013bis}), the wavelengths of the DW radiation peaks can be easily calculated. For our dual core MF, all resonant DW wavelengths are plotted as a function of the soliton wavelength in Fig.~\ref{fig2}. Here the small contribution due to the soliton nonlinear phase shift has been neglected. We verified the validity of this approximation by evaluating the nonlinear contribution to the DW wavelength: for a soliton with a peak power smaller than 10~kW, the longer DW wavelength (upper curve of Fig.~\ref{fig2}) increases by less than 5~nm. 

In the wavelength range considered in Fig.~\ref{fig2} there are always two resonant DWs: for a pump soliton at 1064~nm, the calculated DW wavelengths are 928~nm ($DW_1$, blue curve) and 1587~nm ($DW_2$, green curve). 
A noticeable feature is that the spectral position of  the $DW_1$ spans a window which is two times wider than that of the mate  $DW_2$.

In this work, two different laser sources are used as the pumps to obtain spectral broadening and DW generation in the same 4~m long dual core MF. The first source is a compact microchip Nd:YAG laser (similar to the one used in Ref.~\cite{Wadsworth2004,Stone2008}), whose pulses centered at 1064~nm have a duration of 900~ps.  The input beam was carefully focused on the fiber facet in order to maximize the coupling with the fundamental mode. The second source is a mode-locked fiber laser emitting 120~ps pulses at 1030~nm: its output fiber pigtail was directly spliced to the MF. 
In both cases, the peak power injected inside the MF fiber can be varied up to a maximum value of a few kilowatts.

\begin{figure}[htbp]
\centerline{\includegraphics[width=.8\columnwidth]{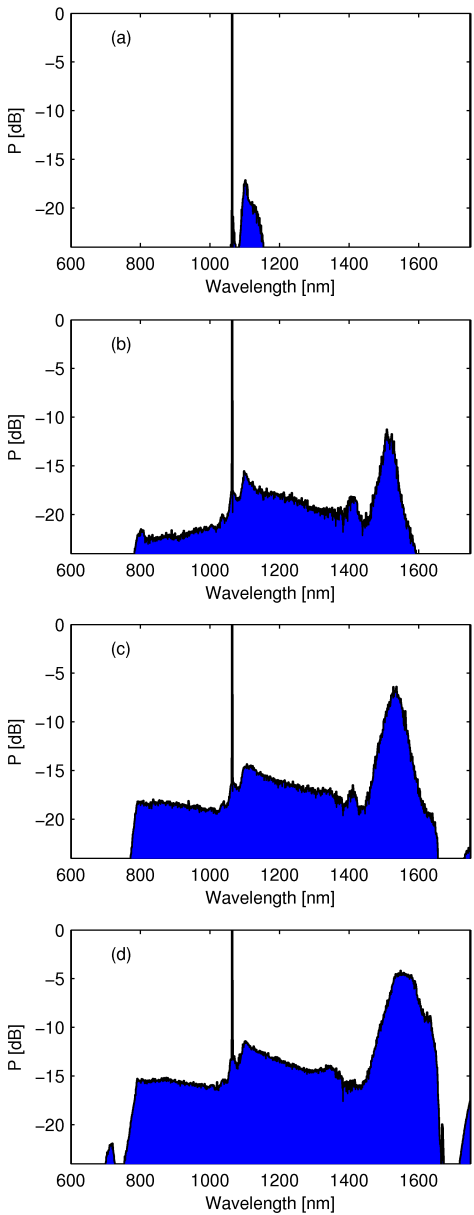}}
\caption{Measured output spectra when the source is the Nd:YAG microchip laser at 1064~nm for four different levels of injected input peak power:(a) 0.1~kW, (b) 0.6~kW, (c) 2~kW, (d) 4~kW.}
\label{fig3}
\end{figure}

The spectra obtained by using the microchip laser at 1064~nm are shown in Fig.~\ref{fig3}, for four different values of the peak power injected inside the MF. At the lowest power level of 0.1~kW, a broad red-shifted spectral peak appears: the maximum of this peak is 17~dB lower than the residual pump, and it is located at about 1100~nm; this provides a first evidence that the broadening mechanism is quite asymmetric around the pump, thus favoring the energy transfer towards longer wavelengths.  Since this spectrum is measured after 4 m of propagation, the two MI peaks are only scarcely discernible. On the other hand, when measuring the spectra after a propagation of less than 1 m the initial growth of two almost symmetric MI peaks is apparent (as in the experimental data reported in Ref.~\cite{Manili2012}). By increasing the input pump power, the output spectrum broadens further, and a DW (henceforth $DW_2$) peak grows around 1510~nm. The peak at 1100~nm is still present, and it is roughly 5~dB higher than the spectral density dip at the long wavelength side of the pump. Note that spectral broadening towards shorter wavelengths abruptly stops at about 800~nm. At the maximum injected pump power of 4~kW,  the $DW_2$ peak is shifted to 1548~nm, with a spectral intensity that is only 4~dB lower than the residual pump peak. Quite strikingly, the $DW_2$ carries an impressive 50\% of the total output average power.  We also underline that the beam image at the fiber output (observed with an infrared camera around 1500~nm) did agree quite well with the profile predicted by the mode solver.

In the measured spectra, there is always a spectral hump around 1100~nm which corresponds to the largest fraction of solitons generated by the input pulse break-up. By considering an average soliton central wavelength of 1100~nm, the $DW_2$ peak is expected to be at 1575~nm (see Fig.~\ref{fig2}), which is reasonably close to the measured value; the residual discrepancy is ascribed to a small error in our estimation of the MF refractive index transverse profile.
The other resonant dispersive wave peak $DW_1$, which is predicted to occur at 874~nm, was not observed in the measured spectra, at least in the form of a sharp and isolated peak.
In fact the spectrum broadens towards shorter wavelengths in a rather uniform way: its short-wavelength edge, located at around 800~nm, is limited by the condition of group-velocity matching with the infrared part of the SC (see Fig.~\ref{fig1}(a)). Hence its location is only slightly affected by variations in the input pump power.
We expect that in our experiments the DWs are shed by a large number of solitons 
with a distribution of peak powers and time widths:  this situation is markedly different from the case of DWs that are emitted by means of stable femtosecond laser sources. 

Solitons generated by the 1064~nm pump do not remain confined around 1100~nm, but travel towards longer wavelengths, and give rise to a long spectral tail in the 1100-1400~nm range (see Fig.~\ref{fig3}(b)): this spectral red-shift is due to TOD, and is enhanced by the Raman effect \cite{Mitschke1986}. The more the pulses are red-shifted, the more they slow down (see Fig.~\ref{fig1}(a)): for this reason, solitons interact and frequently overlap in time. Colliding solitons may significantly enhance the efficiency of DW generation with respect to the single soliton case, see Refs.~\cite{Erkintalo2010,Erkintalo2010bis,Tonello2015}. In our experiments, the large number of collisions among the solitons originating from pump pulse break-up will thus largely contribute to reinforce the energy transfer towards the DWs. 

Note that the soliton tunneling mechanism \cite{Kibler2007,Poletti2008,Guo2013} may also lead to a high DW conversion efficiency. However we may rule out the soliton tunneling mechanism for the generation of $DW_2$. In fact, the $DW_2$ spectral peak is located well inside the normal dispersion region, and it starts growing well before solitons approach the ZDW at the border of the normal dispersion barrier (whose calculated value is 1353 nm).  The initial stages of the spectral broadening process have been reported in Ref.~\cite{Manili2012}. An experimental spectrum similar to the one that we obtain at high powers (Fig.~\ref{fig3}(c)-(d)) was previously reported for a solid core, double zero dispersion wavelength MF by Chapman et al. \cite{Chapman2010}, who measured a broad and intense peak at 1.98~$\mu m$ beyond a 5~dB flat SC,  but in that experiment the DW peak only grew after that the red-shifted solitons had approached the barrier of normal dispersion.

We also observe that our spectra exhibit a hump close to 1400~nm (this is clearly visible in Fig.~\ref{fig3}(b)-(c)), 
which could be due to the accumulation of solitons, if we suppose that the actual ZDW in our MF 
is situated more than 50~nm above the numerically estimated value. 
The growth of a DW in the infrared was also observed by Kudlinski et al. \cite{Kudlinski2009} by using a MF with two zero dispersion wavelengths: however in those experiments the DW was about 15~dB lower than the residual pump, likely because of the low energy or the limited spectrum of the accumulated solitons. 

Once again, it must be underlined that the DW that is observed in Fig.~\ref{fig3} is so intense because it is fueled by the solitons as soon as they are formed from the break-up of the pump pulse. Moreover, these solitons carry a large fraction of the total energy that is coupled inside the MF.
The mechanisms that are responsible for the exceptional growth of $DW_2$ at 1.548~$\mu m$ as well as for the negligible growth of the spectral intensity of $DW_1$ around 0.9~$\mu m$ will be more deeply investigated in the following sections.

\begin{figure}[htbp]
\centerline{\includegraphics[width=.8\columnwidth]{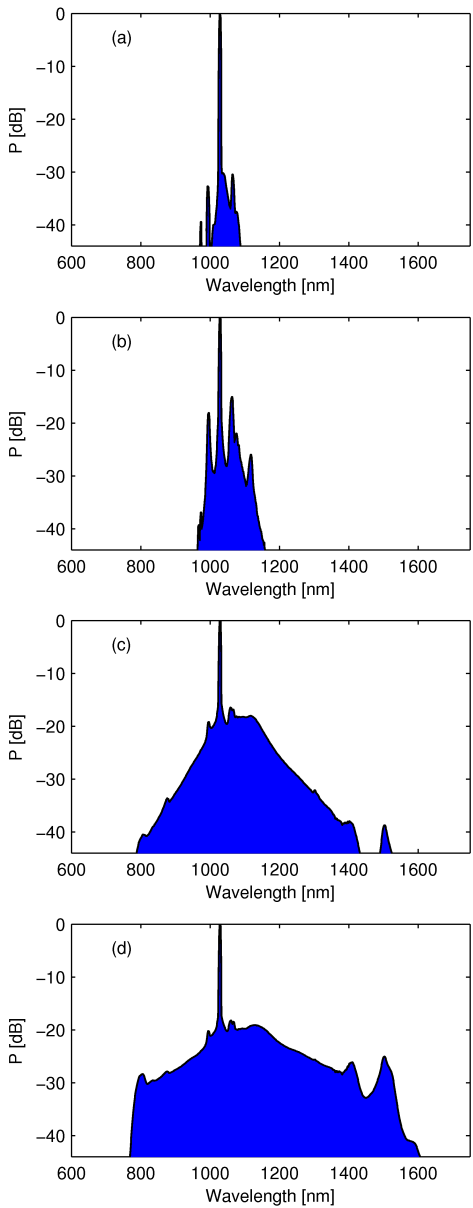}}
\caption{Measured output spectra when the source is the mode-locked fiber laser at 1030~nm for four different levels of injected input peak power:(a) 1.37~kW, (b) 2.7~kW, (c) 5.3~kW, (d) 8~kW.}
\label{fig4}
\end{figure}

For comparison, we report in Fig.~\ref{fig4} the spectra that are measured at the MF output by using the fiber laser emitting at 1030~nm. As can be seen, even at very high input powers a drastic reduction in DW conversion efficiency is observed. At the lowest level of pump power that is shown in Fig.~\ref{fig4}(a), two small MI peaks on both sides of the residual pump can be identified at the wavelengths of 1065~nm and 994~nm, respectively. The spectral density distribution preserves the same qualitative shape even when the peak pump power is increased up to the value of 2.7~kW (Fig.~\ref{fig4}(b)), but an isolated DW peak appears at the wavelength of 1504 nm when the injected peak power reaches the value of 5.3~kW (Fig.~\ref{fig4}(c)). At such power the output spectrum extends from 828~nm to 1418~nm (when considering the -40~dB level). In Fig.~\ref{fig4}(c) the DW peak amplitude is 39~dB lower than the residual pump at 1030~nm, and the DW peak rises up to -25~dB at the maximum available power of 8~kW (Fig.~\ref{fig4}(d)).

The laser operating at 1030~nm has a pulse duration that is about 7 times shorter than that of the microchip laser at 1064~nm.  Although it is reasonable to suppose that a reduction in pulse duration causes a reduction in the spectral broadening \cite{Andreana2012} and hence in the population of solitons, with a consequent drop in the emission of DWs, this cannot entirely explain the dramatic drop of DW generation efficiency which is observed when using the fiber laser. We may thus conclude that
the DW emission efficiency and its spectral position are strongly connected with both the carrier wavelengths of the solitons and the MF dispersion profile.

\section{Spectro-temporal characterization of the DW}

\begin{figure}[htbp]
\centerline{\includegraphics[width=.8\columnwidth]{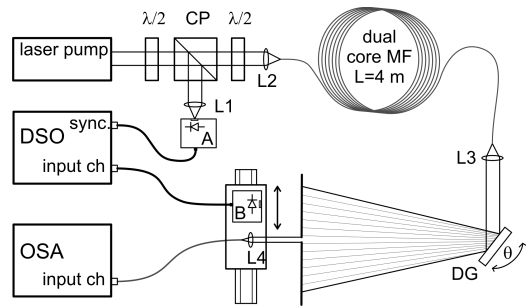}}
\caption{Experimental setup for the spectro-temporal characterization of the DWs emitted by the dual concentric core MF; L1,L2,L3,L4 are micro-lenses and the $\lambda/2$ elements are half-wavelength plates to control the polarization state
(see the text for a detailed description of the setup).}
\label{fig5}
\end{figure}

We limit our spectro-temporal analysis to the case where a maximum conversion efficiency is obtained, {\it i.e.},  
when pumping the MF at 1064~nm. Figure~\ref{fig5} presents the experimental setup that has been used to measure 
the relative time-delay among the different spectral components of the DW.  
A cube polarizer (CP) splits the optical beam of the microchip laser in two directions in order to obtain, by means of a photodiode (A), a trigger signal to be used in an electrical 16~GHz bandwidth digital sampling oscilloscope (DSO). Light is injected into the inner core of the 4 m long MF by means of a micro-lens (L2); the outgoing spectrum is collimated through a micro-lens (L3), dispersed by means of a diffractive grating (DG) and an aperture of 1~mm: this structure works as a tunable bandpass filter with a bandwidth of 6~nm. 
Each spectrum slice is then first measured by an optical spectrum analyzer (OSA) and then by a photodiode (B), which can be  interchanged by means of a linear translation stage. It is thus possible to measure the relative time delay between the arrival time of a given spectrum slice in B and the trigger signal in A. It is worthy to remind that the microchip laser is operating in Q-switching mode, thus the pulse-to-pulse timing jitter needs a self-referencing with the time of emission of each pulse, and this is the role of photodiode A. Both photodiodes A and B are InGaAs PIN diodes, with 12.5~GHz bandwidth.

\begin{figure}[htbp]
\centerline{\includegraphics[width=.8\columnwidth]{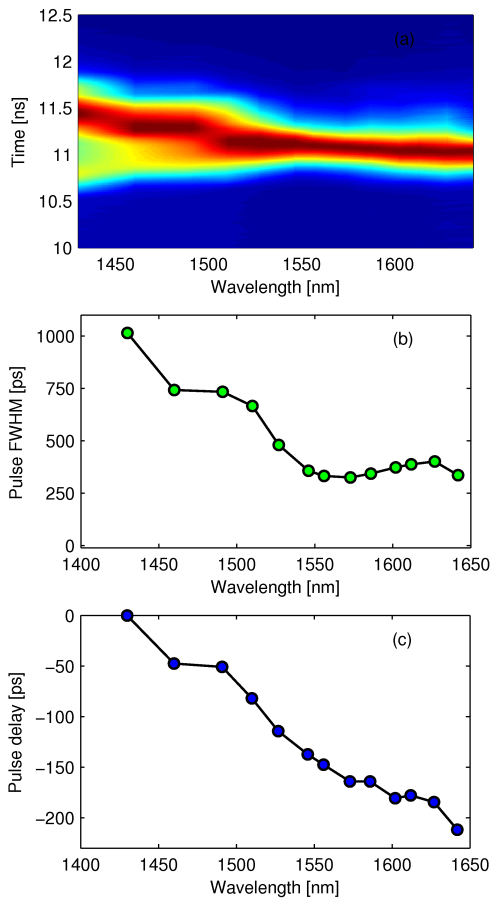}}
\caption{Spectro-temporal experimental characterization of $DW_2$ when the pump is the Nd:YAG laser: (a) spectrogram around the central wavelength of $DW_2$; (b) $DW_2$ pulse FWHM versus wavelength; (c) $DW_2$ pulse delay versus wavelength (the choice of the time origin is arbitrary). The filled circles in (b) and (c) are the measured values and the continuous lines are guides for the eye.}
\label{fig6}
\end{figure}

We show in Fig.~\ref{fig6}(a) a collection of temporal profiles corresponding to different slices of the DW;  panels (b) and (c) of the same figure report summary graphs for the pulse full-width at half-maximum (FWHM) and the pulse delay recorded at different central wavelengths. The spectral region at 1650~nm leads the wavelengths around 1450~nm by 163~ps after 4 m of fiber, which fits fairly well with the 31.5~ps/m $\times$ 4 m=126 ps of time delay predicted by the numerical analysis of the MF guiding properties. However, we underline that the pulse duration that is used in our experiment is longer than the time delay under test. Therefore our conclusions are drawn by estimating the shift of the center of mass of the recorded temporal profiles. 
From our measurements, it is confirmed that the infrared part of the spectrum is present in the leading edge of the pulse, owing to the normal dispersion region of the MF. 
To conclude this section, we may note that the measured time delay remains nearly unchanged for all considered power levels, as the delay is related to the linear guiding properties of the dual concentric core MF.

\section{Intensity of the DW  upon pump wavelength}

In this Section we present a detailed analytical and numerical study of the generation of dispersive waves in our dual-core MF, in order to further clarify the physical mechanisms which may explain the observed large conversion efficiencies and their dependence upon the pump wavelength and MF dispersion profile.
By numerically solving the Generalized Nonlinear Schr\"odinger Equation (GNLSE) by means of the split-step Fourier method, the input pulse break-up and the resulting formation of solitons can be calculated both in the time and in the frequency domains.  In our numerical simulations we did not include the variation of the mode effective area, since it only has a minor contribution on the qualitative dynamics of the propagation of solitons in the region 1064-1400~nm.

Let us first consider for simplicity a long square shaped pulse with peak power P=1.5~kW and width of $T_L$=500~ps: this {\it ansatz} could represent
the internal part of a sub-nanosecond laser pulse. If we consider a central wavelength of 1064~nm, the numerically calculated group velocity
dispersion is $\beta_2=-5.88\times 10^{-27}$ ~s$^2$/m. By assuming a nonlinear coefficient  $\gamma=21\times 10^{-3}$~W$^{-1}$m$^{-1}$,
we may estimate a MI frequency $f_{MI}=16.4$~THz, hence a modulation period $T_{MI}=61$~fs.
From the ratio $T_L/T_{MI}$, we expect to obtain about $8000$ solitons, with individual energy equal to $E_S=P/f_{MI}=91$~pJ.
When considering only first-order solitons, we may calculate a corresponding reference time duration parameter $T_0=2|\beta_2|/(\gamma E_S)=6$~fs, which corresponds to a soliton FWHM of $10.6$~fs, and peak power $P_S=|A_S|^2=E_S/(2T_0)=7.7$~kW.  
Although this reasoning is only approximate (the pulse widths are of the order of just a few optical cycles), 
the previous values may provide a first-order estimation of the magnitude and number of solitons that are generated from the decay of the initial long pump pulse.

From our full numerical solutions of the GNLSE  with either single sub-nanosecond input pump pulses or with trains of femtosecond solitons (not shown here), we may conclude that the formation of the observed huge DW peak cannot be explained if the input solitons are individually considered. In other words, the experimentally observed DW peak cannot be reproduced by simply adding all the DWs that are generated by each single soliton. Actually, when numerically simulating the formation and evolution of a bunch of these solitons along the dual concentric core MF, it is apparent that they reach different peak amplitudes and central wavelengths. Hence, due to their different group velocities, solitons interact and undergo multiple collisions.
Indeed it was already pointed out that soliton-soliton collisions may increase the DW intensity by orders of magnitude \cite{Erkintalo2010,Erkintalo2010bis,Tonello2015}. Thus we focus our analysis here on the role of the soliton parameters in determining the relative amplitude of the two predicted DWs: this problem is relevant since only one peak (namely, $DW_2$) is clearly visible in our measured spectra.

As discussed in the previous Section, for an input pump pulse centered at 1064~nm, during the first stage of propagation the generated solitons have wavelengths that are mostly located around 1100~nm  (Fig.~\ref{fig3}(a)). As the measured spectra show, at the end of the 4~m long MF these solitons entirely fill the spectral region where dispersion is anomalous (Fig.~\ref{fig3}(b)-(d)). Numerical simulations confirm this picture, where solitons slow down and red-shift; the Raman effect gives a contribution to this spectral shift, however our numerical results show that a considerable red-shift is present even if the Raman coefficient is set to zero. 

\begin{figure}[htbp]
\centerline{\includegraphics[width=.8\columnwidth]{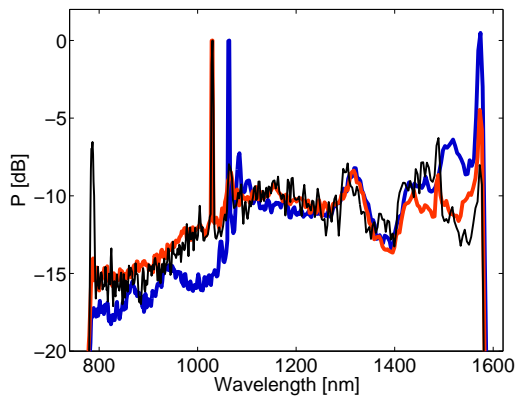}}
\caption{  (Color online) Numerically calculated output spectra (the Raman effect is included) with different input pulses with a common peak power of 1~kW. 
Blue thick curve: pulse duration of 200~ps and central wavelength of 1064~nm; 
red thick curve: pulse duration of 200~ps and central wavelength of 1030~nm; 
black thin curve: pulse duration of 30~ps and central wavelength of 1030~nm. }
\label{fig7}
\end{figure}

 The blue thick curve of Fig.~\ref{fig7} shows the numerically computed spectrum at the output of the 4~m long double core MF, computed by assuming an input pump pulse at 1064~nm, with a duration of 200~ps and a peak power of 1~kW.  The Raman effect was included in our simulations. This spectrum can be compared with the measured values in Fig.~\ref{fig3}(d): both the numerical result and the experiment show a residual MI peak around 1100~nm, an intense $DW_2$ and a region of wide spectral broadening in the 800-1600~nm range (see also the numerical results reported in Ref.~\cite{Manili2012}). The red thick curve in Fig.~\ref{fig7} is obtained by shifting the carrier wavelength of the input pump pulse to 1030~nm. The output spectrum is qualitatively similar to the previous one, it exhibits a decrease in $DW_2$, but it does not agree with the experiment of Fig.~\ref{fig4}. In fact, the pulses emitted by the fiber laser at 1030~nm are seven times shorter than the pulses emitted by the microchip laser at 1064~nm. We thus decreased the pump pulse duration to 30~ps: the black thin curve of Fig.~\ref{fig7} shows a consistent drop in $DW_2$ in satisfactory agreement with the experimental data of Fig.~\ref{fig4}(d).
Additionally, note that the black curve also exhibits a sharp sideband peak at 788~nm which is generated by
four-wave mixing in the early stages of the propagation (the associated sideband is at 1486~nm), i.e., before the pulse break-up.  Fig.~\ref{fig4}(d) shows the presence of a little sideband on the short wavelength limit of the SC,
and a similar feature is also weakly visible in Fig.~\ref{fig3}(b).

  The nanosecond pump pulse may decay in a significant number of solitons, nevertheless we focus our attention on the propagation of just one of them, since 
we may expect that the soliton central wavelength has an impact not only on the DW position, 
but also on its amplitude.
  
\begin{figure}[htbp]
\centerline{\includegraphics[width=.8\columnwidth]{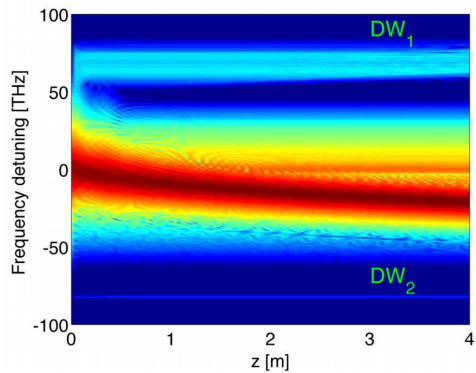}}
\caption{ Numerically calculated spectral evolution of an input soliton of 30 fs centered at 1100 nm.  $DW_1$ and $DW_2$
are both clearly identifiable.}
\label{fig8}
\end{figure}

  As a first example, we show in Fig.~\ref{fig8} the spectral evolution of a 30~fs soliton 
initially centred at 1100~nm. The color code represents the spectral intensity in logarithmic scale, and again we included in these simulations the Raman effect. 
Both dispersive waves $DW_1$ and $DW_2$ are clearly evident, and
the soliton exhibits a marginal red-shift. Most of the radiated energy is carried by $DW_1$
in agreement with the theory that we will illustrate later on, however this situation is far from 
what we observed in the experiments.

\begin{figure}[htbp]
\centerline{\includegraphics[width=.8\columnwidth]{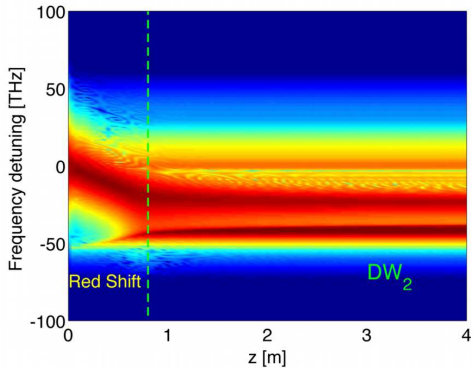}}
\caption{  Numerically calculated spectral evolution of an input soliton of 30 fs centered at 1200 nm.
The soliton follows a rapid red-shift and then $DW_2$ is emitted.}
\label{fig9}
\end{figure}

Figure~\ref{fig9} shows the evolution of the same 30~fs soliton but with a central wavelength of 1200~nm. Now the soliton and DW generation dynamics is completely different from the previous case, and it can be decomposed in two steps: a first stage leading to a net red-shift of about 
30~THz is followed by a relevant emission of $DW_2$ that inhibits further red-shift.
Surprisingly, such a huge soliton self-frequency shift induces a change in the intensity of
$DW_2$, while leaving its spectral position nearly unchanged.

\begin{figure}[htbp]
\centerline{\includegraphics[width=.8\columnwidth]{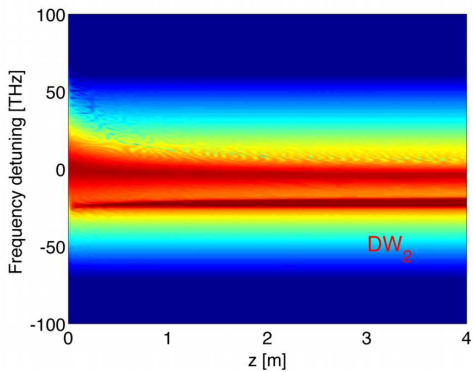}}
\caption{ Numerically calculated spectral evolution of an input soliton of 30 fs centered at 1300 nm.
There is no noticeable soliton red-shift and the $DW_2$ is emitted since the beginning.}

\label{fig10}
\end{figure}

Larger energy leakages into DWs can substantially reduce the soliton red-shift:  this is the case of
Fig.~\ref{fig10}, where the soliton was initially centered at 1300~nm.  
Figures~\ref{fig9} and \ref{fig10}  show the asymmetry in favour of $DW_2$ which is observed in our experiments. More specifically, Fig.~\ref{fig9} shows how this effect is intensified after the initial stage of red-shift of the soliton central wavelength.

From the measured spectra of Fig.~\ref{fig3} we may expect that the pump pulse break-up eventually leads to a significant population of red-shifted solitons between 1100~nm and about 1400~nm. Simulations, such as those of Fig.~\ref{fig9}, prove that this particular spectral distribution of  solitons contributes to generate a red-shifted $DW_2$ that is much larger than the blue-shifted $DW_1$.

 It would be interesting to extrapolate the dependence of the DW amplitude upon the soliton wavelength, in order to understand the amount of $DW_2$ that one can potentially generate.
To this end, we applied the theory developed by Akhmediev and Karlsson \cite{Akhmediev1995} to the specific case of our MF fiber. 
We evaluated the frequencies of resonant coupling and the initial DW spectrum for different values of the soliton carrier wavelength.  To simplify the approach we do not
take into account the soliton self-frequency shifts that originate from either the Raman effect or the high-order dispersion terms. Still we do consider the full dispersion profile in order to properly localize the dispersive waves. In fact Fig.~\ref{fig9} shows
how, in a first approximation, red-shift and resonant emission can be considered separately.

Let us start from the GNLSE (with Raman effect now neglected) written in the frequency domain as
\begin{equation}
\frac{\partial \hat A}{\partial z}-i\kappa(\omega)\hat A -i\gamma{\cal F}[|A|^2A]=0
\label{gnlse}
\end{equation}
where $A(t,z)$ is the complex envelope of the optical field and $\hat A(\omega,z)={\cal F}[A]=\int_{-\infty}^{+\infty}A(t,z)\exp(i\omega t) \,dt$ is its Fourier transform.
Equation~(\ref{gnlse}) is written in a reference frame moving at the group velocity of the reference physical angular frequency 
$\omega_0$. In what follows we set the reference frequency $\omega_0$ 
as the carrier angular 
frequency of a soliton, so that $\omega$ measures the shift of the angular frequency of the optical field $\omega+\omega_0$ from $\omega_0$. The full dispersion profile of the MF is described by the function $\kappa(\omega)=\beta(\omega+\omega_0)-\beta(\omega_0)
-\beta_{1}(\omega_0) \,\omega$, where $\beta_1(\omega_0)$ has the physical meaning of the reciprocal of the group velocity at $\omega_0$.
Note that varying the reference angular frequencies $\omega_0$ leads to different dispersion profiles $\kappa(\omega)$.

The relative time delay between different spectral components of the field and the reference frame is proportional to
\begin{equation}
k'(\omega)=\frac{\partial \kappa}{\partial\omega}=\beta_{1}(\omega+\omega_0)
-\beta_{1}(\omega_0)
\label{gvm}
\end{equation}
From Eq.~(\ref{gvm}) we can see that $k'(\omega)$ measures the group velocity mismatch (or difference in group delays) between the waves at $\omega+\omega_0$ and the reference frame at $\omega_0$. 

In this work, we mainly focus our attention on solitons that are generated by the long pulse break-up, so it is reasonable to represent the linear operator of the GNLSE as 
$\kappa(\omega)=\beta_{2}(\omega_0) \omega^2 /2+\epsilon \hat H(\omega)$,
where $\beta_{2}(\omega_0)$ is the group velocity dispersion measured at the reference frequency, $\hat H(\omega)$ describes higher-order dispersion terms, and $\epsilon$ is a small parameter that permits to separate short and long length scales in the solution of Eq.~(\ref{gnlse}).

We may thus study the solution of Eq.~(\ref{gnlse}) by using the {\it ansatz} provided by a first order soliton plus an additional small perturbation 
$\hat A(\omega,z)=\hat A_0(\omega)\exp[i\kappa_Sz]+\epsilon F(\omega,z)$. In our notation, $\hat A_0(\omega)$ is the first order soliton solution, $\kappa_S=\gamma |A_S|^2 /2$ is the nonlinear contribution to the soliton wavenumber, and $F(\omega,z)$ is the perturbation. Our {\it ansatz} neglects the soliton frequency shift that may be induced by $\hat H(\omega)$, hence its validity
will be  limited to the early stages of the DW emission process (details on the soliton dynamics in presence of third order dispersion can be found in Refs.~\cite{Wabnitz1994,Akhmediev1995,Gaeta2002,Mussot2010}). 
Now, since the soliton is the first order solution (i.e., with $\epsilon=0$) of Eq.~(\ref{gnlse}), 
we may collect all terms proportional to $\epsilon$ and obtain a linear forced equation for the perturbation $F(\omega,z)$
\begin{equation}
\frac{\partial F}{\partial z}-i\kappa(\omega) F=i \hat H(\omega) \hat A(\omega)\exp[i\kappa_Sz]
\label{lsa}
\end{equation}
The solution of Eq.~(\ref{lsa}), with zero initial condition, can be written as
\begin{equation}
F(\omega,z)=\frac{\hat H(\omega) \hat A_0(\omega)}{\kappa_S -\kappa(\omega)}\left[e^{i\kappa_Sz} -e^{i\kappa(\omega)z}\right]
\label{lsa:solution}
\end{equation} 
Equation~(\ref{lsa:solution}) expresses the well-known property that DWs are fed by the spectrum of the soliton, and that their amplitude is enhanced at those frequencies $\omega_R$ which satisfy the resonant condition $\kappa(\omega_R)=\kappa_S$ \cite{Akhmediev1995}.

When dealing with a large number of unequal solitons, an important issue is the sensitivity of the resonant condition with respect to variations of the soliton peak power $P_S$. 
If we perform a series expansion around the resonance frequency $\omega_R$, we obtain 
$\kappa(\omega_R+d\omega_R)\simeq  \kappa(\omega_R)+d\omega_R\kappa'(\omega_R)$ and, by applying a standard error
analysis for a variation $dP_S$ of the soliton peak power, we may calculate the sensitivity of the resonant frequency to a small change in soliton power with respect to a reference value
\begin{equation}
\frac{d\omega_R}{dP_S}=\frac{\gamma}{2\kappa'(\omega_R)}
\label{sensitivityP}
\end{equation}   
Equation~(\ref{sensitivityP}) shows that the variation of the resonant frequency resulting from a small variation of the soliton power is determined by the group velocity mismatch $\kappa'(\omega_R)=\beta_1(\omega_R+\omega_0)-\beta_1(\omega_0)$ at the resonant frequency $\omega_R+\omega_0$.
Large values of the group velocity mismatch, which bring long temporal delays between the DW and the soliton, will thus result in a small sensitivity of $\omega_R$ to changes of the soliton peak power. 

Equation~(\ref{sensitivityP}) also describes an important property of the generated DWs: soliton amplitude jitter may lead to fluctuations in the DW frequency, because a change in peak power results in a variation of the zeros of the phase-matching condition. Note that a similar change is also observed if a red-shifting soliton is subject to an amplitude reshaping. 
Since DW frequency fluctuations are inversely proportional to $\kappa'(\omega_R)$, in the presence of a distribution of soliton powers, large values of 
$\kappa'(\omega_R)$ will drastically reduce the standard deviation of the DW frequencies. This may explain why the $DW_2$ peak at 1548~nm is narrow-band ($\kappa'(\omega_R)$ has a large modulus there), while there is virtually no peak but only a flat plateau around 800-950~nm for $DW_1$.

A similar conclusion can also be drawn for the sensitivity of the resonance frequency as a result of changes in the soliton
central wavelength. From a close inspection of Fig.~\ref{fig2}, it is clear that the slope of the curve 
for $DW_1$ (lower blue curve) around 1064~nm is more than twice the slope of the curve for $DW_2$ (upper green curve). Numerical simulations (not shown here) attest that small fluctuations of the soliton wavelength indeed lead to a much larger spreading for the $DW_1$ resonance than for the $DW_2$ resonance.
  
\begin{figure}[htbp]
\centerline{\includegraphics[width=.8\columnwidth]{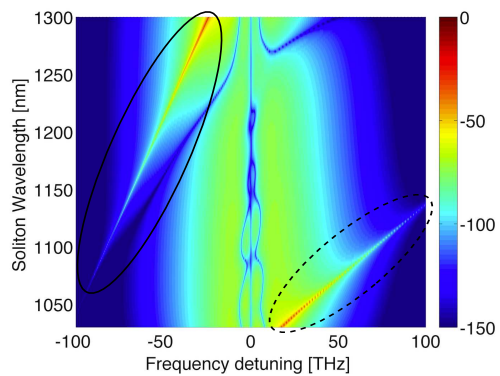}}
\caption{Output normalized spectra (in logarithmic scale) as given by Eq.~(\ref{lsa:solution}) versus the frequency detuning $\omega/2 \pi$ for different central wavelengths of the input first order soliton having a FWHM of 30~fs and for 4 m of propagation. The region where $DW_1$ ($DW_2$) is generated is highlighted by a dashed line (solid line) ellipse.}
\label{fig11}
\end{figure}

In Fig.~\ref{fig11} we illustrate the spectrum of the perturbation as it is predicted from Eq.~(\ref{lsa:solution}), for different pump soliton wavelengths, and a propagation length of 4 m: the generation of $DW_1$, $DW_2$ and of the red-shifted spectral tail (between 1100 nm and 1400 nm) are clearly recognizable. Owing to the perturbative nature of $F(\omega,z)$, the spectrum of Fig.~\ref{fig11} provides an estimate of the output spectrum at all frequencies, except for values close to the soliton reference frequency. This analysis confirms that for soliton carrier wavelengths larger than about 1150~nm the intensity  of the $DW_2$ peak grows larger than that associated with $DW_1$.
The DW amplitude depends not only on the soliton wavelength, but also on its FWHM (which is equal to 30 fs in the simulations of Fig.~\ref{fig11}): this is not surprising since the soliton spectrum appears as a multiplying factor in Eq.~(\ref{lsa:solution}). 

\begin{figure}[htbp]
\centerline{\includegraphics[width=.8\columnwidth]{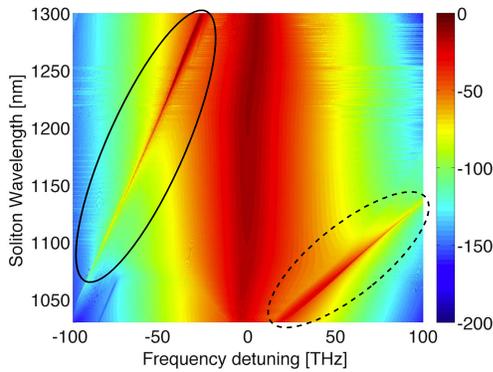}}
\caption{Output normalized spectra (in logarithmic scale) versus the frequency detuning $\omega/2 \pi$ for different central wavelengths of the input first order soliton (FWHM=30~fs) calculated by numerically solving the GNLSE for 4~m of propagation. The region where $DW_1$ ($DW_2$) is generated is highlighted by a dashed line (solid line) ellipse.}
\label{fig12}
\end{figure}

Since higher-order terms were neglected in deriving Eq.~(\ref{lsa:solution}), we have verified its validity by computing the output spectrum through the numerical solution of the GNLSE (see Fig.~\ref{fig12}). The numerically computed DW spectra exhibit a good agreement with the theoretically predicted spectra of Fig.~\ref{fig11}. Thus we may conclude that the approximate analytical solution of Eq.~(\ref{lsa:solution}) provides a quick first estimate of the DW spectrum, which is useful to explain its dependence on the soliton parameters and the fiber dispersion profile. Nevertheless, numerical simulations are still necessary for an accurate evaluation of the DW amplitude, as well as the precise shape of the output spectrum.
 By moving along the different soliton wavelengths of Fig.~\ref{fig11} or Fig.~\ref{fig12} we can see how a red-shift 
of the soliton central wavelength can gradually intensify $DW_2$. 

\section{Conclusions}
We presented an extensive experimental and numerical study of the highly efficient generation of a DW spectral peak at telecom wavelengths in a dual-core microstructured optical fiber, pumped by a near infrared microchip laser. Both experiments and simulations agree in their predictions that the conversion efficiency into the DW spectrum strongly depends on the value of the pump wavelength, in combination with the dispersion profile of the fiber. Moreover, in spite of the fact that the phase-matching condition between solitons and DWs provides multiple resonances, we found that the conversion efficiency is by far more efficient for the DW that is closest to the pump solitons. We further presented an approximate but analytical expression for the DW spectrum which explicitly contains the dispersion profile of the fiber, and thus provides an useful tool for estimating the DW position and amplitude. Moreover, the intensity of the analytical DW spectrum is proportional to the amplitude of the soliton spectral tail at the resonance wavelength, which confirms the observation that stronger DWs are generated when resonance occurs closer to the center wavelength of the pumping soliton ensemble. 

We believe that our results can help in understanding the process of highly efficient DW generation in dispersion-engineered optical fibers. In particular, our analysis may be used as a guideline for the optimization and reverse engineering of specialty fibers with the purpose of emitting a target and high energy DW peak in any desired spectral region of interest for a particular application. 

S.W. is also with Istituto Nazionale di Ottica of the Consiglio Nazionale delle Ricerche. We acknowledge the partial support from the  R\'egion Limousin (C409-SPARC), and by the Italian Ministry of University and Research (MIUR, Project No. 2012BFNWZ2)

\end{document}